\documentclass[aps,prc,preprint,showpacs,floatfix,nofootinbib]{revtex4}
\usepackage{amssymb}
\usepackage{amsmath}
\usepackage{graphicx}
\usepackage{bm}
\usepackage{color}

\newcommand{\be}{\begin{equation}}
\newcommand{\ee}{\end{equation}}
\newcommand{\bel}[1]{\be\label{#1}}
\newcommand{\re}[1]{Eq.~(\ref{#1})}
\newcommand{\ds}{\displaystyle}
\newcommand{\hsp}{\hspace*{1pt}}
\newcommand{\hspm}{\hspace*{.5pt}}

\begin{document}

\title{Equation of state and sound velocity of hadronic gas\\ with hard-core interaction}

\author{L.M. Satarov$^{1,2}$, K.A. Bugaev$^{1,3}$, and I.N. Mishustin$^{1,2}$}

\affiliation{$^1$\mbox{Frankfurt Institute for Advanced Studies, D-60438 Frankfurt am Main, Germany}\\
$^2$National Research Center ''Kurchatov Institute'', 123182 Moscow, Russia\\
$^3$Bogolyubov Institute for Theoretical Physics, 03680 Kiev, Ukraine}

\begin{abstract}
Thermodynamic properties of hot and dense hadronic systems with a hard-sphere interaction are calculated in the
Boltzmann approximation. Two parametrizations of pressure as a function of density are considered: the first one,
used in the excluded volume model and the second one, suggested earlier by Carnahan and Starling. The results are given
for one--component systems containing only nucleons or pions, as well as for chemically equilibrated mixtures of pions,
nucleons and delta resonances. It is shown that the Carnahan-Starling approach  can be used in a much broader range
of hadronic densities as compared to the excluded volume model. In this case superluminal sound velocities appear only
at very high densities, in the region where the deconfinement effects should be already important.
\end{abstract}


\maketitle

\section{Introduction}

Properties of strongly interacting matter under extreme conditions attract a great interest
of researches working in several fields including relativistic heavy--ion collisions, physics of compact
stars and the early universe. In recent years a significant progress was achieved in lattice calculations of
the equation of state (EoS)
at high temperatures and low baryon densities. However, the lattice approach cannot be used reliably at low
temperatures and high baryon densities. The information on the EoS in this domain remains a subject of model
calculations. It is obvious that realistic calculations
of the EoS of dense hadronic systems should take into account a strong interaction between hadrons.

A hard--sphere interaction (HS\hsp I) is one of the most popular methods to implement short range repulsion effects
for calculating thermodynamic properties of multiparticle systems. In this approach
the particles of a sort $i$ are represented by hard spheres of the radius $R_{\hsp i}$.
It is assumed that particle move freely unless the distance $r_{ij}$ between centers of any pair $i,j$ becomes
equal to $R_{\hsp i}+R_j$.
It is postulated that the potential energy of $ij$-interaction is infinite at smaller $r_{ij}$\hsp. Originally such
an approximation has been suggested by Van-der-Waals~\cite{V1910} to describe properties of dense gases and liquids.
Later on the HS\hsp I--based models were successfully used by many authors in condensed matter physics~\cite{Han06,Mul08}.
A similar approach, the so-called
excluded volume model~(EVM), has been applied in~\cite{Cle83,Ris91,Sat09,Bug08,Ste11,Bug13,Oli13} to describe
the EoS of hot and dense hadronic matter. These studies revealed a very strong sensitivity of the EoS to parameters
of the short-range repulsion between hadrons. In particular, it has been shown in~\cite{Ris91,Sat09} that a reasonable
phase diagram of strongly interacting matter can only be obtained after accounting for finite sizes of hadrons.
Our present study is aimed at a more realistic description of the HS\hsp I effects in the hadronic EoS.

Unfortunately, the Van-der-Waals approach is essentially nonrelativistic. As a consequence, it can not be safely
applied when the sound velocity of matter $c_s$ becomes comparable with the light velocity.
It is well known that the EVM violates the casuality con\-di\-tion\hspm\footnote
{
Units $\hbar=c=1$ are used throughout the paper.
}
~$c_s<1$ at high enough baryon densities~\cite{Sat09}. Attempts to remove this drawback were made in~Refs.~\cite{Kag93,Bug08}
(see also~\cite{Sin96,Gor12}). Moreover, it will be shown below that the EVM becomes inaccurate at
high densities when the total volume of constituents exceeds 10--20\%
of the system volume. By comparing with the virial expansion~\cite{Han06,Lan80} one may conclude that
this model overestimates the contribution of non-binary interactions to pressure.

On the other hand, numerical simulations of one-component liquids with HS\hsp I show~\cite{Han06} that the
Carnahan-Starling approximation (CSA) of pressure~\cite{Car69} successfully works up to much higher densities then
in the EVM. Below we use the CSA and~EVM to calculate properties of thermodynamically equilibrated hadronic systems
containing mesons and baryons. For simplicity, we include into consideration only the lightest nonstrange hadrons
(pions and nucleons) as well as the lightest baryonic resonance $\Delta$ (with the mass $m_\Delta=1232~\textrm{MeV}$)
in the zero-width approximation\hsp\footnote
{
 We also neglect the isospin and Coulomb effects as well as possible creation of baryon--antibaryon pairs.
}.
An important feature of such systems is that partial numbers of different species~$N_i$ where $i=\pi, N, \Delta\ldots$
are, in general, not conserved due to presence of inelastic processes and resonance decays. These numbers are not
inde\-pen\-dent and should be determined from the conditions of chemical equilibrium~\cite{Sat09}:
\bel{cheq}
\mu_{\hspm\pi}=0\hsp,\,\,\mu_N=\mu_\Delta=\ldots=\mu_B\hsp .
\ee
Here $\mu_{\hspm i}$ is the chemical potential of the $i$-th species and $\mu_B$ is the baryon chemical potential. At
given total baryon number $B$, system volume~$V$ and temperature $T$ one can determine~$\mu_B$ from the relation
$\sum\limits_i{N_i}b_{\hsp i}=B$
where~$b_{\hsp i}$ is the baryon charge of the $i$-th species.

Up to now the information about properties of multi--component systems with HS\hsp I is rather scarce~\cite{Mul08}.
In this paper we consider several representative cases: first, we study the $N+\Delta$
and \mbox{$\pi+N+\Delta$} mixtures with equal sizes of all hadrons and second, the \mbox{$\pi+N+\Delta$} system
assuming that baryons have equal radii, $R_\Delta=R_N$, and pions are point-like, $R_{\hsp\pi}=0$\hsp\footnote
{
Note that small~\cite{Bug13} or even vanishing~\cite{Oli13} pion radii are favored by recent fits of hadron multiplicities
observed in central heavy--ion collisions at the AGS, SPS and RHIC bombarding energies.
}.
The main emphasis is given to calculating the sound velocity. According to our analysis, the CSA predicts a much softer EoS,
with
smaller \mbox{$c_s$--\,values} than in the EVM. Choosing reasonable values of hadronic radii, we show that acasual states
in the CSA are shifted to baryon densities $n_B\gtrsim 1~\textrm{fm}^{-3}$. It is expected that at such densities the
deconfinement effects, in particular, the formation of a quark--gluon phase should be already important.

The paper is organized as follows: in Sec.~II we consider one-component systems with hard sphere particles. First we
introduce parametrizations of pressure in the EVM and CSA. Then we calculate properties of an ideal gas in the Boltzmann
approximation.
Analytic expressions for shifts of thermodynamic functions due to HS\hsp I are obtained in Sec.~II\hspm C and~IID.
The EoS and sound velocities of nucleon and pion matter are analyzed in Sec.~IID and~IIE\hspm .
In Sec.~III we study properties of the \mbox{$N+\Delta$} and \mbox{$\pi+N+\Delta$} mixtures. The summary and outlook
are given
in Sec. IV. In the Appendix we derive a general formula for the sound velocity of a~rela\-tivistic gas.

\section{One--component hadronic systems}

\subsection{Compressibility and virial expansion\label{comp}}
In this section we consider a monodisperse system containing only one sort of hard-sphere particles with radius~$R$.
Below we disregard the effects of Fermi or Bose statistics i.e. all calculations are done in the classical (Boltzmann)
approximation.  In this case one can write the following expression for pressure as a function of temperature and particle
density~$n=N/V$~\cite{Han06,Mul08}:
\bel{pres1}
P=n\hsp TZ(n)=P_{\hsp\rm id}Z(n)\hsp .
\ee
Here $P_{\hsp\rm id}$ is the ideal gas pressure and $Z$ is the ''compressibility'' factor, which depends only on the
dimensionless ''packing'' fraction $\eta=n\hsp v$ where $v=4\pi R^{\hsp 3}/3$ is the proper volume of a single particle.
At small $\eta$ one can use a universal virial expansion~\cite{Han06}
\bel{vire}
Z=1+4\hspm\eta+10\hsp\eta^2+\ldots
\ee
This expansion is not applicable\hsp\footnote
{
The most dense state of the considered systems corresponds to the ordered (face--centered cubic) lattice
with \mbox{$\eta=\frac{\pi}{3\sqrt{2}}\simeq 0.74$}\hsp . Direct Monte-Carlo simulations show~\cite{Mul08} that
the liquid-solid phase transition in a~one--component matter with HS\hspm I occurs in the interval $0.49<\eta<0.55$\hsp.
}
for $\eta$ exceeding about 0.5\hsp . Equation (\ref{vire}) may be applied to estimate the accuracy of EoS calculations
for multiparticle systems with HS\hsp I.

Instead of (\ref{vire}), different~ana\-ly\-tical approximations for $Z$ are used by many authors.
For example, the following Van-der-Waals--motivated parametrization is used in the~EVM:
\bel{zevm}
Z_{\hsp\rm EVM}=\frac{1}{1-4\hsp\eta}~.
\ee
One can see that such an ansatz leads to inaccurate results at high enough $\eta$\hspm . Indeed, comparison
of the r.h.s of (\ref{zevm}), decomposed in powers of $\eta$\hsp , with \re{vire} shows that only first two
terms of the virial expansion are correctly reproduced in the EVM. It is clear that densities~\mbox{$n>0.25/v$}
can not be reached in this model due to the divergence of pressure
at~\mbox{$\eta=0.25$}\hsp . As demonstrated in Ref.~\cite{Sat09}, the EVM leads to superluminal sound velocities
already at $\eta\gtrsim 0.2$\hsp . This is a consequence of a too stiff density dependence of pressure assumed in this model.

     \begin{figure*}[hbt!]
          \centerline{\includegraphics[trim=0 7.5cm 0 8.5cm, clip, width=0.8\textwidth]{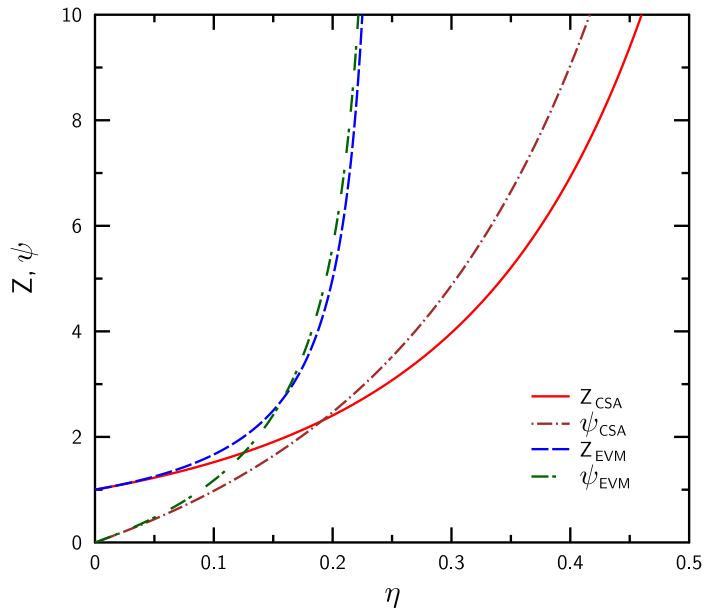}}
        \caption{(Color online)
         Compressibility factor $Z$ and function $\psi$ (see~\re{psin}) for different values of packing
	  fraction $\eta$ calculated within the EVM and CSA.}
        \label{fig1}
        \end{figure*}
On the other hand, the Carnahan--Starling parametrization~\cite{Car69}:
\bel{zcsa}
Z_{\rm CSA}=\frac{1+\eta+\eta^2-\eta^3}{(1-\eta)^3}
\ee
is able to reproduce rather accurately~\cite{Han06} the first eight terms of the virial expansion for~$Z(n)$\hspm .
It agrees well with numerical calculations at $\eta\lesssim 0.5$\, i.e.~up to the boundary of liquid phase. Note that
both above--mentioned parametrizations give similar results in the region $\eta\ll 1$
where~\mbox{$Z\simeq 1+4\hsp\eta$}. This is illustrated in Fig.~\ref{fig1}. One can see that at $\eta\gtrsim 0.2$
the Carnahan-Starling EoS is indeed noticeably softer as compared with the EVM.

Equation (\ref{pres1}) gives pressure as a function of canonical variables: temperature $T$ and density $n$.
As explained above, in the situation when particle densities are not fixed, more appropriate
variables are temperature and chemical potential $\mu$. It is possible to calculate other thermodynamic functions,
in particular the energy and entropy densities, $\varepsilon$ and $s$, if the dependence \mbox{$\mu=\mu\hsp (T,n)$}
is known. To get explicit expressions for these functions, it is convenient to calculate first the free energy
density $f=\mu\hsp n-P$ as a function of $T$ and $n$\hspm . Then one can use thermodynamical identities~\cite{Lan80}
\bel{ther1}
\varepsilon=f+T\hsp s,~~~~s=-\left(\frac{\partial f}{\partial T}\right)_n.
\ee

\subsection{Thermodynamic functions of ideal gas}

Let us start from calculating thermodynamic functions of an ideal gas of particles with the
mass $m$ and the spin-isospin degeneracy factor $g$. In the Boltzmann approximation one can write
down~\cite{Sat09} the equation relating the particle density and the chemical potential
\bel{denmid}
n=\phi\hsp (\hsp T)\hsp\exp{\left(\frac{\mu_{\hsp\rm id}}{T}\right)},\,\,\phi\hsp (T)
\equiv\frac{g\hsp m^3}{2\pi^2}\hsp\frac{K_2(x)}{x}\,.
\ee
Here $x=m/T$, $K_n(x)$ is the McDonald function of $n$-th order and the subscript 'id' implies the ideal gas limit.
The function $\phi$ has the meaning of the ideal gas density in the case of zero chemical potential.

From \re{denmid} and formulae of preceding section we get the following expressions for thermodynamic functions of
the ideal gas:
\begin{eqnarray}
&&\mu_{\hsp\rm id}=T\ln{\frac{n}{\phi\hsp (\hsp T)}}\,,\label{cpid}\\
&&f_{\hsp\rm id}=\mu_{\hsp\rm id}\hsp n-P_{\hsp\rm id}=
n\hsp T\left[\hsp\ln{\frac{n}{\phi\hsp (T)}}-1\right],\label{feid}\\
&&s_{\hsp\rm id}=n\left[\hsp\ln{\frac{\phi\hsp (\hsp T)}{n}}+\xi\hsp (T)\right],\label{entid}\\
&&\varepsilon_{\hsp\rm id}=n\hsp T{\bigl[}\hsp\xi\hsp (T)-1{\bigr]}\,,\label{endid}
\end{eqnarray}
where
\bel{ksif}
\hspace*{-1.4cm}\xi\hsp (T)=T\frac{\phi^{\,\prime}(T)}{\phi\hsp (T)}+1=x\hsp\frac{K_3(x)}{K_2(x)}\,.
\ee
Unless otherwise stated, we denote by prime the derivative with respect to $T$.
According to \re{endid}, in the ideal gas limit, the heat capacity per particle
\mbox{$\widetilde{C}=n^{-1}\left(\frac{\ds\partial\hsp\varepsilon}{\ds\partial\hsp T}\right)_n$} is a function
of temperature only:
\bel{hcap}
\widetilde{C}=\left[\hsp T\hsp(\hsp\xi-1)\hsp\right]^{\,\prime}=x^2+3\,\xi-(\hsp\xi-1)^{\,2}\,.
\ee

The sound velocity is an important characteristics of EoS which gives the propagation speed of small
density perturbations in the matter rest frame. In absence of dissipation the adiabatic sound velo\-city
squared is equal
to~\cite{Lan87}
\bel{adsv}
c_s^{\hsp 2}=\left(\frac{\partial P}{\partial\hsp\varepsilon}\right)_{\hspace*{-1pt}\sigma}\,,
\ee
where the subscript $\sigma$ in the r.h.s. means that the derivative is taken along the Poisson adiabat,
i.e. at constant entropy per particle\hsp\footnote
{
In a general case, when particle numbers are not conserved $\sigma$ equals the entropy per baryon.
}:
$\sigma=s/n=\textrm{const}.$
One can rewrite~\re{adsv} in the form
\bel{adsv1}
c_s^{\hsp 2}=\frac{(\partial P/\partial\hsp n)_{\hsp T}+
(\partial P/\partial\hsp T)_{\hsp n} (\partial\hsp T/\partial\hsp n)_{\hsp\sigma}}
{(\partial\hsp\varepsilon/\partial\hsp n)_{\hsp T}+
(\partial\hsp\varepsilon/\partial\hsp T)_{\hsp n} (\partial\hsp T/\partial\hsp n)_{\hsp\sigma}}\,.
\ee

Using~Eqs.~(\ref{entid}), (\ref{hcap}) and the relation
\mbox{$d\sigma=d\hspace*{0.5pt}n\hsp\partial\hspace*{0.5pt}\sigma/\partial n +
d\hspace*{0.5pt}T\hsp\partial\hspace*{0.5pt}\sigma/\partial T=0$} we get
in the ideal gas limit
\bel{adtdn}
n\left(\frac{\partial\hsp T}{\partial\hsp n}\right)_{\sigma}=\frac{T}{\widetilde{C}}\,\,.
\ee
After calculating the derivatives of $P,\varepsilon$ in (\ref{adsv1})~and using~\re{adtdn}
we obtain the following formula for the sound velocity of the monodisperse ideal gas:
\bel{csid}
c_s^{\,\rm id}=\sqrt{\xi^{-1}\left(1+\widetilde{C}^{-1}\right)}\,.
\ee
One can see that the sound velocity of the classical ideal gas is a function of tempe\-ra\-ture~only.

In the nonrelativistic limit, $T\ll m$, using the asymptotic formulas for McDonald functions,
one gets the approximate expressions
\bel{asrel}
\xi\simeq x+\frac{5}{2}+\frac{15}{8\hsp x}+\ldots\hsp ,~~~~\widetilde{C}\simeq\frac{3}{2}+
\frac{15}{4\hsp x}-\frac{15}{2\hsp x^2}+\ldots
\ee
Substituting (\ref{asrel}) into (\ref{csid}), we get the well--known non-relativistic expression
$c_s^{\,\rm id}\simeq\sqrt{\frac{5T}{3\hsp m}}$ for the sound velocity of a monoatomic ideal gas.

In the opposite, high temperature limit, $T\gg m$, one obtains from Eqs.~(\ref{ksif})--(\ref{hcap})
\bel{urrel}
\xi\simeq 4+\frac{x^2}{2}+\ldots\hsp ,~~~~\widetilde{C}\simeq 3-\frac{x^2}{2}+\ldots\,.
\ee
 This leads to the ultrarelativistic result $c_s^{\,\rm id}\simeq 1/\sqrt{3}\simeq 0.577$\hsp. One can show that
 $c_s^{\,\rm id} (T)$ is a~monotonically increasing function with the asymptotic value $1/\sqrt{3}$\,.

\subsection{Contribution of HS\hsp I}\label{chsi}

In this section we consider deviations from the ideal gas limit for particles with HS\hsp I\hspm .
Let us denote by $\Delta A$ the shift of any quantity from its ideal gas value:
\bel{exvn}
\Delta A\equiv A-A_{\hsp\rm id}\,.
\ee
It is clear that $\Delta A\to 0$ in the dilute gas limit $n\to 0$\hsp . Integrating  the thermodynamic
relation~\cite{Lan80} $d\mu=\frac{\ds 1}{\ds n}\hsp (dP-s\hspm d\hspm T)$ along the density axis (at fixed $T$)\hsp ,
one obtains the equation
\bel{excp1}
\Delta\mu\hsp (T,n)=\int\limits_0^{n}\hsp\frac{d\hspm n_1}{n_1}\frac{\partial\Delta P\hsp (T,n_1)}{\partial\hsp n_1}~.
\ee
Here we have used the condition $\lim\limits_{n\to\hsp 0}{\Delta\mu}=0$. Substituting $\Delta P=n\hsp T(Z-1)$,
one arrives at the relation $\Delta\mu=T\psi (n)$ where
\bel{psin}
\psi (n)=Z(n)-1+\int\limits_0^{n}\hsp\frac{d\hspm n_1}{n_1}\left[\hsp Z(\hsp n_1)-1\hsp\right].
\ee
The same formula for $\Delta\mu$ has been obtained earlier in
Ref.~\cite{Mul99}. Using further~\re{cpid} we finally get the equation for the chemical potential
$\mu=\mu_{\hsp\rm id}+\Delta\mu$  as a function of $T$ and $n$:
\bel{chpt2}
\mu=T\left[\ln{\frac{n}{\phi\hsp (T)}}+\psi(n)\right].
\ee

By solving (\ref{chpt2}) with respect to $n$ and substituting the result
into \re{pres1} one can calculate pressure as a function of grand-canonical variables $T,\mu$\hsp . In particular,
this may be useful for finding possible phase transitions by using the Gibbs construction. Parametrizations of the
compressibility factor introduced in Sec.~\ref{comp} are rather useful because they permit an analytical
integration in~\re{psin}. For example, in the EVM
\mbox{Eqs.~(\ref{pres1}), (\ref{zevm}), (\ref{psin})} give the following result
\bel{psevm}
\psi_{\hsp\rm EVM}=Z-1+\ln{Z}=\frac{bP}{T}+\ln{\left(1+\frac{bP}{T}\right)}\,,
\ee
where $b=4\hspace*{0.5pt}v$ is the ''excluded volume'' introduced by Van-der-Waals.
Substituting (\ref{psevm}) into~(\ref{chpt2}) leads to a simple formula for baryon chemical potential
\bel{bcp3}
\mu=T\ln{\frac{P}{T\phi\hsp (T)}}+bP~~~~~~~~~~(\textrm{EVM})\,.
\ee
One can regard~\re{bcp3} as the implicit equation for $P=P\hsp (T,\mu)$. In the considered
case solving~\re{bcp3} with respect to $P$ is equivalent to solving~\re{chpt2} with respect to~$n$.
It is worth noting that in the EVM the shift of chemical potential from the ideal gas value~($b=0$)
is linear in pressure. But this conclusion is not universal: it does not hold in the CSA.

Indeed, substituting~(\ref{zcsa}) into~\re{psin} gives the following formula
\bel{pscsa}
\psi_{\hspm\rm CSA}=\frac{3-\eta}{(1-\eta)^{\hsp 3}}-3\,.
\ee
In Fig.~\ref{fig1} we present numerical values of $\psi (n)$ in the EVM and CSA. One can see that
at given $T,n$  the values of $\psi$ and, therefore, deviations of chemical potential
from the ideal gas values are  larger in the~EVM\hsp .

\subsection{Nucleonic matter}\label{chsi1}

Let us consider first a system consisting of nucleons ($m=939~\textrm{MeV}\hspace*{-1pt},~g=4$).
In this case $n$ is the conserved baryon density, which together with temperature defines the thermodynamic state.
At fixed~$n$ and $T$ the shift of
free energy density due to nucleon interactions equals \mbox{$\Delta f=n\hsp\Delta\mu-\Delta P$}.
Using further~Eqs. (\ref{feid}) and (\ref{psin}) we obtain the expression for the free energy density of
interacting nucleons
\bel{feint}
f=n\hsp T\left\{\hsp\ln{\frac{n}{\phi\hsp (\hsp T)}}-1+\int\limits_0^{n}\hsp\frac{\ds dn_1}{\ds n_1}
\left[Z(\hsp n_1)-1\right]\right\}.
\ee
Equations (\ref{ther1}), (\ref{feint}) lead to the following formulae for entropy and energy densities:
\begin{eqnarray}
&&s=n\left\{\hsp\ln{\frac{\phi\hsp (\hsp T)}{n}}+\xi\hspm
(\hspm T)-\int\limits_0^{n}\hsp\frac{\ds dn_1}{\ds n_1}\left[Z(\hsp n_1)-1\right]\right\},\label{entd1}\\
&&\varepsilon=f+Ts=n\hsp T\left[\hsp\xi (\hspm T)-1\right],\label{end1}
\end{eqnarray}
where $\xi\hspm (\hspm T)$ is defined in~\re{ksif}\hspm .

     \begin{figure*}[hbt!]
          \centerline{\includegraphics[trim=0 7.5cm 0 8.5cm, clip, width=0.8\textwidth]{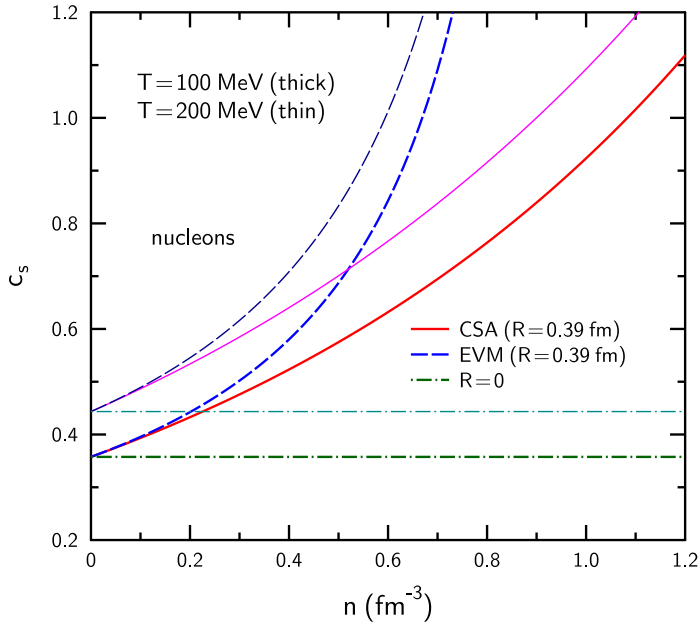}}
        \caption{(Color online)
         Sound velocity of nucleon gas as a function of density for two values of
         temperature $T=100$ (thick lines) and $200$ (thin lines) MeV. The solid and dashed curves are calculated
	  by using compressibility functions proposed in the CSA and EVM. The dash--dotted lines are obtained
         in the limit of point--like nucleons ($R=0$)\hspm .}
        \label{fig2}
        \end{figure*}
From Eqs.~(\ref{endid}) and (\ref{end1}) one can see  that HS\hsp I does not
produce any shift of the energy density as compared to the ideal gas of point--like nucleons\,\footnote
{
This result is rather obvious. It is clear that the energy per particle, $\varepsilon/n$\hspm , for one--component
systems with classical hard--\hsp sphere particles should depend only on temperature, at least for densities below
the liquid--\hsp solid transition. Therefore, increasing the density at fixed $T$ does not change $\varepsilon/n$\hspm .
}.
As a consequence, the isochoric heat capacity
\mbox{$C_{\rm V}=(\partial\hsp\varepsilon/\hsp\partial\hspace*{1.5pt}T)_{\hsp n}$} is the same as in the
ideal gas:
$C_{\rm V}=n\,\widetilde{C}(T)$ where $\widetilde{C}(T)$ is given by~\re{hcap}.
According to~(\ref{entd1}), the entropy per particle $\sigma=s/n$ is reduced due to hard-core interaction of nucleons.
This leads to a modification of the Poisson adiabat~(\mbox{$\sigma=\textrm{const}$}) in the~\mbox{$n-T$} plane as
compared to the ideal gas. Indeed, using~\re{entd1}, we obtain for the isentropic process
\bel{adtdn1}
n\left(\frac{\partial\hsp T}{\partial\hsp n}\right)_{\sigma}=Z\hsp T\hsp\widetilde{C}^{\hsp -1}.
\ee
Comparing this result with~\re{adtdn} we conclude that the slope of the Poisson adiabat
of nucleonic matter increases with density due to the appearance of the compressibility factor~$Z>1$.

     \begin{figure*}[hbt!]
          \centerline{\includegraphics[trim=0 7.5cm 0 8.5cm, clip, width=0.8\textwidth]{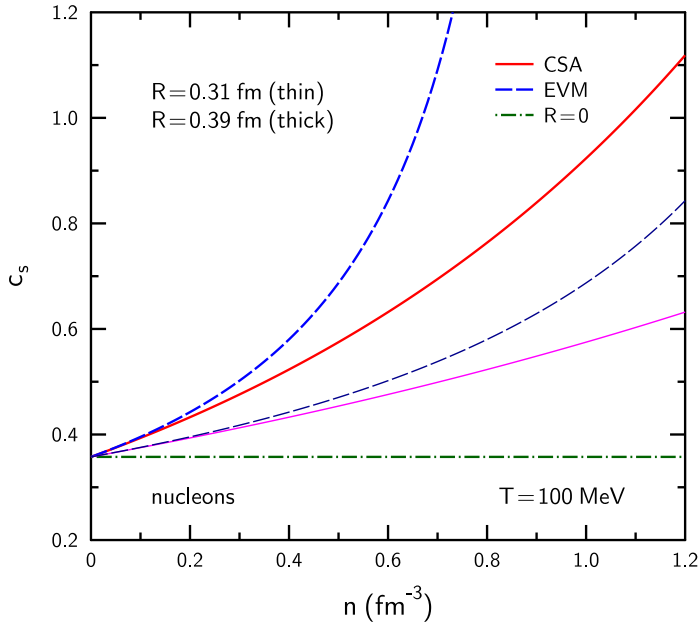}}
        \caption{(Color online)
         Sound velocity of nucleon matter as a function of density for different
         values of parameter $R$\hsp . The dashed and solid curves are calculated, respectively, in the EVM and CSA.
         The dash--dotted line corresponds to point--like nucleons.}
        \label{fig3}
        \end{figure*}
The sound velocity can be obtained from~Eqs.~(\ref{pres1}), (\ref{adsv1}), (\ref{end1})--(\ref{adtdn1})\hspm .
This leads to the analytic expression
\bel{adsvn}
c_s^{\hsp 2}=\frac{1}{\xi+Z-1}\left[(nZ)^{\hsp\prime}+Z^{\hsp 2}\widetilde{C}^{\hsp -1}\right],
\ee
where prime means the derivative with respect to $n$\hspm .
In the ideal gas limit $Z\to 1$ this formula coincides with~\re{csid}. In the case of a nucleon
gas at realistic tempe\-ra\-tures~\mbox{$T\ll m$}, using~the relations (\ref{asrel}) one can derive
the approximate formula
\bel{adsvn1}
c_s^{\hsp 2}\simeq\frac{(nZ)^{\hsp\prime}+2Z^{\hsp 2}/3}{Z+x+3/2}~,
\ee
where $x=m/T$. Equa\-tions~\mbox{(\ref{adsvn})--(\ref{adsvn1})} clearly show that HS\hspm I leads to superluminal
sound velocities, $c_s>1$, at high enough densities where $Z$ is large
(for further discussion, see~\cite{Sat09}). This is especially evident
in the~EVM where $(nZ)^{\hsp\prime}=Z^{\,2}$. According to~\re{adsvn1}, in this case $c_s^{\hsp 2}$ is proportional to $Z$
at large $Z$.

     \begin{figure*}[hbt!]
          \centerline{\includegraphics[trim=0 5.8cm 0 5cm, clip, width=1.2\textwidth]{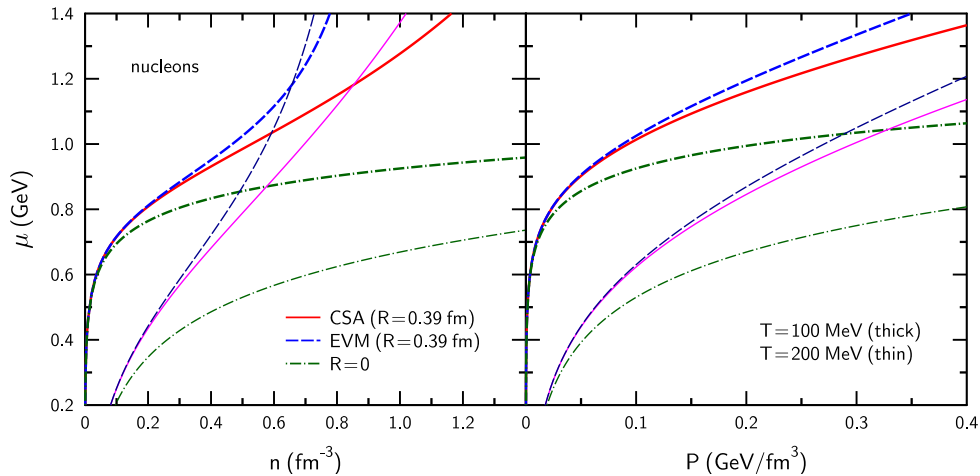}}
        \caption{(Color online)
         Chemical potential of nucleon gas as a function of density (left panel) and
         pressure (right panel) calculated in the CSA (solid lines) and EVM (dashed lines)
         at tempera\-tures~100 and 200 MeV. The dashed--dotted lines correspond to point--like nucleons.}
        \label{fig4}
        \end{figure*}
In Fig.~\ref{fig2} we compare the sound velocities values, calculated by using the
parametrizations~(\ref{zevm}) and (\ref{zcsa}) for two typical values of temperature.
We have chosen the nucleon radius~\mbox{$R=0.39~\textrm{fm}$} which
corresponds to the excluded volume $b=1~\textrm{fm}^3$, used previously in
Ref.~\cite{Sat09}\hsp . Again one can see that the~CSA predicts a much softer EoS
(i.e.~smaller~$c_s$) than the~EVM. Our calculations show, that at realistic temperatures
$T\lesssim 200~\textrm{MeV}$ the sound velocity in the~CSA remains below unity
up to rather large densities $n\simeq 0.9~\textrm{fm}^{-3}$. On the other hand,
superluminal sound velocities appear in the EVM at much smaller $n$\hspm .
According to~Fig.~\ref{fig2}, deviations from the ideal gas limit $R\to 0$ become significant
already at subnuclear densities~$n\sim 0.1~\textrm{fm}^{-3}$.

Figure~\ref{fig3} demonstrates that the sound velocity is very sensitive to the choice of the
particle size $R$\hspm . Note that a 20\%
reduction of $R$\hspm , from 0.39 to 0.31 fm, corresponds to the two--fold decrease of the excluded
volume $b$\hspm . It is seen that the difference between CSA and EVM is smaller for lower $R$\hspm .
Figure~\ref{fig4} shows the results for the baryon chemical potential as a~function of nucleon density
and pressure. One can see that at $n\gtrsim 0.4~\textrm{fm}^{-3}$ the CSA indeed predicts significantly
smaller values of~$\mu$ as compared to the EVM. On the other hand, at given $\mu$ the
pressure in the CSA is noticeably larger than in the EVM. This makes the nucleon phase more stable at high
densities as compared to the EVM.

\subsection{Pion matter}

Let us consider now thermodynamic properties of matter composed of finite--\,size \mbox{''thermal''} pions
with the vacuum mass $m_{\hspm\pi}=140~\textrm{MeV}$ and the statistical weight $g_{\hspm\pi}=3$. As before,
we assume the hard--sphere interaction of particles and perform all calculations in the Boltzmann approximation.
To emphasize specific features of the pion system we introduce the subscript '$\pi$'. Using
Eqs.~(\ref{pres1}), (\ref{chpt2}), one can write the following equations for pressure and chemical potential of pions
\bel{prmup}
P_\pi=T\hsp n_{\hspm\pi} Z(n_{\hspm\pi}),~~~\mu_{\hsp\pi}=T\left[\ln{\frac{n_{\hspm\pi}}
{\phi_{\hsp\pi}\hspm (T)}}+\psi(n_{\hspm\pi})\right],
\ee
where $\phi_{\hsp\pi}$ and $\psi$ are defined in (\ref{denmid}) and (\ref{psin}) (with
$m=m_{\hspm\pi}, g=g_{\hspm\pi}$).

     \begin{figure*}[hbt!]
          \centerline{\includegraphics[trim=0 7.5cm 0 8.5cm, clip, width=0.8\textwidth]{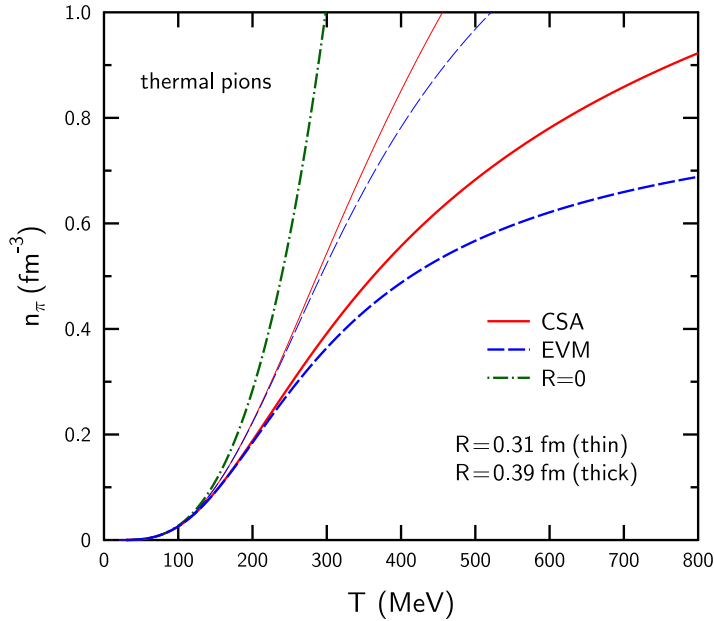}}
        \caption{(Color online)
         Equilibrium density of pions as a function of temperature for different values of parameter $R$.
         The dashed and solid curves are obtained, respectively, in the EVM and CSA. The dash--dotted line
         corresponds to ideal gas of point--like pions.}
        \label{fig5}
        \end{figure*}
At fixed temperature one can find equilibrium values of density $n_{\hspm\pi}=n_{\hspm\pi} (\hsp T)$ and
other thermodynamic functions from the condition of chemical equilibrium $\mu_{\hspm\pi}=0$\hsp .
Then we obtain the following implicit equation for density of pions
\bel{eqpd}
n_{\hspm\pi}=\phi_{\hsp\pi}\hsp (T)\,e^{-\hsp\ds\psi (n_{\hspm\pi})}.
\ee
As one can see from~\re{eqpd}, finite size effects suppress the pion density as compared
to the ideal gas limit $\psi\to 0$\hsp . This is illustrated in Fig.~\ref{fig5} where
we compare the results of the EVM and CSA, for several values of the hadronic radius $R$\hsp . One can see
 noticeable deviations from the ideal gas already at $T\simeq 150~\textrm{MeV}$, but a significant difference between
 the CSA and EVM calculations appears only at unrealistically high temperatures $T\gtrsim~400~\textrm{MeV}$.
 Such a behavior follows from a relatively slow increase of pion packing ratio with temperature. Note that
 short--range repulsive interactions of pions should also suppress possible Bose--enhancement effects at high
 temperatures.

 One can easily calculate the entropy density of interacting pion gas. In the case $\mu_{\hspm\pi}=0$\hspm , using
 the thermodynamic relation $s_{\hsp\pi}=dP_\pi/dT$, one has
\bel{entdpi}
s_{\hsp\pi}=T\hsp\frac{d\hsp n_{\hspm\pi}}{d\hsp T}\,(n_{\hspm\pi} Z\hsp)^{\hsp\prime}+n_{\hspm\pi}
Z=n_{\hspm\pi}\hsp (Z+\xi_\pi-1)\,,
\ee
where prime denotes the derivative with respect to the density $n_{\hspm\pi}$ and $\xi_\pi$ is defined in~\re{ksif}\hsp .
In the second equality we use the relation
\bel{pdder}
T\frac{d\hsp n_{\hspm\pi}}{d\hsp T}=\frac{n_{\hspm\pi}\hsp (\xi_\pi-1)}{1+n_{\hspm\pi}\psi^{\hsp\prime}}=
\frac{n_{\hspm\pi}\hsp (\xi_\pi-1)}{(n_{\hspm\pi} Z\hsp)^{\hsp\prime}}
\ee
which follows from~\re{eqpd} after taking the derivative with respect to $T$. According to
Eqs.~(\ref{entdpi}), (\ref{urrel}) the entropy per pion
$s_{\hspm\pi}/n_{\hspm\pi}$ equals approximately $Z+3$ at $T\gg m_\pi$\hsp . This value exceeds the corresponding
ratio for massless point--like pions ($Z=1$).

Equation~(\ref{entdpi}) leads to the following formula for the energy density
\mbox{$\varepsilon_{\pi}=Ts_{\hsp\pi}-P_\pi$}:
\bel{endpi}
\varepsilon_{\pi}=n_{\hsp\pi} T\hsp (\xi_\pi-1)\,.
\ee
From this result one can see that at given temperature the energy per particle is the same as in the ideal pion gas.
Using (\ref{pdder})--(\ref{endpi}), (\ref{hcap}) we get the equation for the heat capacity per
pion~\mbox{$\widetilde{C}_\pi=n_{\hspm\pi}^{-1}d\hsp\varepsilon_{\pi}/d\hsp T$}:
\bel{hcpp}
\widetilde{C}_\pi=x_\pi^2+3\hsp\xi_\pi+(\xi_\pi-1)^2\left[\frac{1}
{(n_{\hspm\pi} Z\hsp)^{\hsp\prime}}-1\right],
\ee
where $x_\pi=m_{\hspm\pi}/T$.

     \begin{figure*}[hbt!]
          \centerline{\includegraphics[trim=0 7.5cm 0 8cm, clip, width=0.8\textwidth]{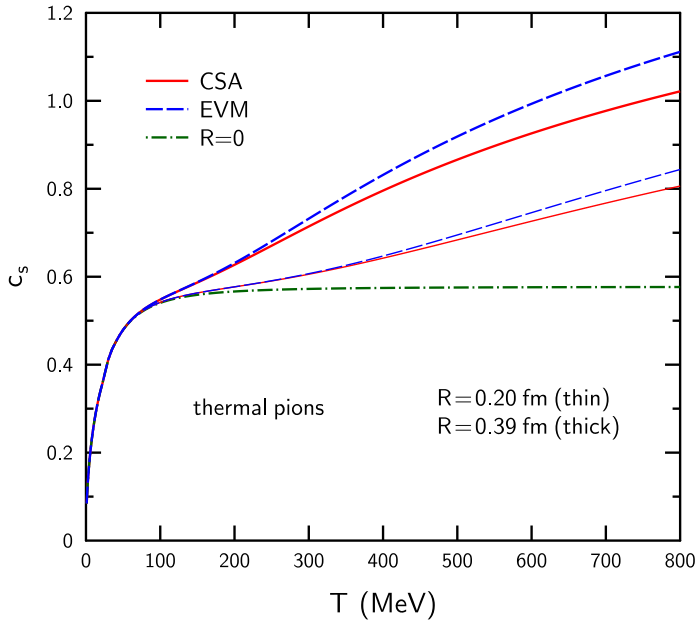}}
        \caption{(Color online)
        Sound velocity of pion gas as a function of temperature for different values of parameter $R$.
         The dashed and solid curves are calculated, respectively, in the EVM and CSA. The dash--dotted line
	  corresponds to point--like pions.}
        \label{fig6}
        \end{figure*}
Finally we obtain the following formula for the sound velocity squared
\bel{svpg}
c_s^{\hsp 2}=\frac{dP_\pi}{d\hsp\varepsilon_\pi}=\frac{s_{\hsp\pi}}{n_{\hspm\pi}
\widetilde{C}_\pi}=\frac{Z+\xi_\pi-1}{\widetilde{C}_\pi}\,.
\ee
In the ideal gas limit $Z\to 1$ one gets $c_s=(3+x_\pi^2/\xi_\pi)^{-1/2}=(3+x_\pi\hsp K_2/K_3)^{-1/2}$.
By using~(\ref{urrel}) we arrive at the approximate relation
\bel{csurr}
c_s^{\hsp 2}\simeq\frac{1}{3}\,(Z+n_{\hspm\pi} Z^{\hsp\prime}\hsp)\hspace*{-1pt}\left(1+
\frac{n_{\hspm\pi} Z^{\hsp\prime}}{Z+3}\right)^{-1},
\ee
in the ultrarelativistic case $x_\pi\ll 1$\hsp. Figure~\ref{fig6} shows the results of
$c_s$--\,calculations with the parameters $R=0.20~\textrm{and}~0.39~\textrm{fm}$\hsp . One can see that
at~$T\gtrsim 200~\textrm{MeV}$ the obtained sound velocities noticeably exceed the asymptotic ideal gas
value $c_s=1/\sqrt{3}$. The calculations show, that these velocities become superluminal only at unrealistically
high temperatures at which hadrons
should melt~\cite{Aok09,Baz12}.

\section{Hadronic mixtures}
\subsection{General remarks}

Let us consider now a multi-component hadronic matter composed of particles of different
kinds $i=1,2\ldots$ Most detailed information about the EoS of this matter can be obtained
if one knows its pressure $P=P\hsp (\hspm T,n_1,n_2\ldots)$ as a function of temperature $T$ and partial densities
$n_{\hspm i}=N_i/V$\hspace*{-2pt}. As before, we neglect the quantum effects and assume that particles interact
via HS\hspm I. In this case one can write down~\cite{Mul08} first two terms of the virial expansion of pressure
in powers of $n_{\hspm i}$:
\bel{vird1}
\frac{P}{nT}=1+\sum\limits_{i,j}b_{\hsp ij}\hsp x_i\hsp x_j+\ldots
\ee
Here $n=\sum\limits_i n_i$ is the total density, $x_i=n_i/n$ and coefficients
$b_{\hsp ij}=\frac{\ds 2\hspm\pi n}{\ds 3}(R_{\hsp i}+R_{\hspm j})^{\hsp 3}$, where~$R_{\hsp i}$ is the radius
of the $i$-th species. If particle radii are the same ($R_{\hsp i}=R$ for all $i$) the second term in the r.h.s. equals
$4\hspm\eta$ where \mbox{$\eta=4\pi\hspace*{-1.2pt}R^{\,3}n/3$}\hspm . In this limit most of the results for
one-- and multi--component systems will be formally the same. In particular, one may use~\re{pres1} and the
formulae for thermodynamic functions from Sec.~\ref{chsi1}
by identifying the variable~$n$ with the total density of all species.

It is possible to calculate the shift of the free energy density, $\Delta f$,
for any multi--component system if one knows its pressure as a function of temperature and partial densities.
Below we use the method suggested in Ref.~\cite{Ald55}. Using the thermodynamic
relation \mbox{$dF=-PdV$} for the change of total free energy $F$ in the isothermal process, one can write down
the equation connecting the shifts of $F$ and $P$:
\bel{dfdp}
\Delta F=\int\limits_V^\infty d\hsp V_*\,\Delta P\left(T,\frac{N_1}{V_*},\frac{N_2}{V_*}\ldots\right).
\ee
The r.h.s. of this equation is equal to the work done by particle interactions during the isothermal
com\-pres\-sion of matter from an asymptotically large volume to $V_*=V$\hspace*{-2pt}. Introducing the variable
$\alpha=V/V_*$ one obtains the expression for $\Delta f=\Delta F/V$:
\bel{shfe}
\Delta f=\int\limits_0^1 \frac{d\hsp\alpha}{\alpha^2}\,\Delta P\hspm(\hsp T,\alpha\hsp n_1,\alpha\hsp n_2\ldots)\,.
\ee
For a one--component matter with HSI, substituting $\Delta P=n\hspm T\hspm (Z-1)$, we return to the formulae,
obtained in Sec.~\ref{chsi1}.

It is easy to derive exact results for mixtures where one of the components consists of point--like particles.
Namely, let us consider a two--component system where the ratio of particle radii~$R_{\hspm 2}/R_1$ is
small. In the limit $R_{\hspm 2}\to 0$ one can regard the component $i=2$ as an ideal gas but in the reduced
''free'' volume $\widetilde{V}=V-N_1v_1=V\hsp (1-\eta_{\hspm 1})$. Here $v_1$ and $\eta_{\hspm 1}$ are, respectively,
the proper volume and the packing fraction of particles $i=1$. The partial pressure of the first component may be
written analogously to~\re{pres1}. This leads to the following equation for pressure of a~two-component
mixture with $R_{\hspm 2}/R_1\ll 1$~\cite{Mul08}
\bel{pr2cs}
P(\hspm T,n_1,n_2)=n_1TZ(n_1)+\frac{n_2T}{1-\eta_1}\,.
\ee
The last term is the partial pressure of the second component $P_2=\widetilde{n}_2T$.
Here~\mbox{$\widetilde{n}_2=N_2/\widetilde{V}$} is the ''local'' density of particles $i=2$ which is larger than
the ''average'' density \mbox{$n_2=N_2/V$}\hspace*{-2pt}. Using \re{pr2cs} one can easily prove the validity of the
virial theorem (\ref{vird1}) in the limit~\mbox{$n_1,n_2\to 0$}\hspm . In fact, instead of particles $i=1$  we can
consider an arbitrary multi-com\-po\-nent mixture composed of hadrons with the same radii. In this
case $n_1$ equals the total density of such a~mixture.

\subsection{The $N+\Delta$ matter}\label{ndm}

In this section we consider the EoS of a chemically equilibrated binary mixture of baryons:
nucleons ($N$) and the lightest $\Delta$ resonances
($m_\Delta=1232~\textrm{MeV}\hspace*{-1.5pt},\hsp g_\Delta=16$)\hspm . This system is characterized
by two canonical variables: temperature and the baryon
den\-sity~\mbox{$n_B=n_N+n_\Delta$}. We assume that all baryons have the same radii, i.e.
$R_{\hspm N}=R_{\hsp\Delta}=R$\hsp . In this case HS\hspm I does not distinguish~$N$
and $\Delta$, therefore, the hadronic pressure can be written as~$P=n_B\hspm T\hspace*{-1pt}Z(n_B)$\hspm . Same arguments as used
in deriving~\re{chpt2} lead to the equation for chemical potential of the~\mbox{$i$-th} species ($i=N,\Delta$):
\bel{chpt3}
\mu_{\hspm i}=T\left[\ln{\frac{n_{\hsp i}}{\phi_{\hsp i}\hspm (T)}}+\psi(n_B)\right].
\ee
Here $\phi_{\hsp i}$ is defined in \re{denmid} with the replacement
$m\to m_{\hsp i}, g\to g_{\hspm i}$.

From the condition of chemical equilibrium $\mu_N=\mu_\Delta=\mu_B$ we get the expressions
\begin{eqnarray}
&&\mu_B=T\left[\ln{\frac{n_{\hspm B}}{\phi_N+\phi_\Delta}}+\psi(n_B)\right],\label{chemp}\\
&&n_{\hsp i}=n_{\hspace*{-1pt}B}\hsp w_{\hspm i}(T),~~~w_\Delta=1-w_N=\frac{\phi_\Delta}
{\phi_N+\phi_\Delta}\,.\label{pden1}
\end{eqnarray}
The relative fractions of $i$-th baryons, $w_{\hspm i}$, depend on temperature only and coincide with
corresponding values for the ideal gas of $N+\Delta$ baryons~\cite{Gal79}\hspm .

     \begin{figure*}[hbt!]
          \centerline{\includegraphics[trim=0 6cm 0 5cm, clip, width=1.2\textwidth]{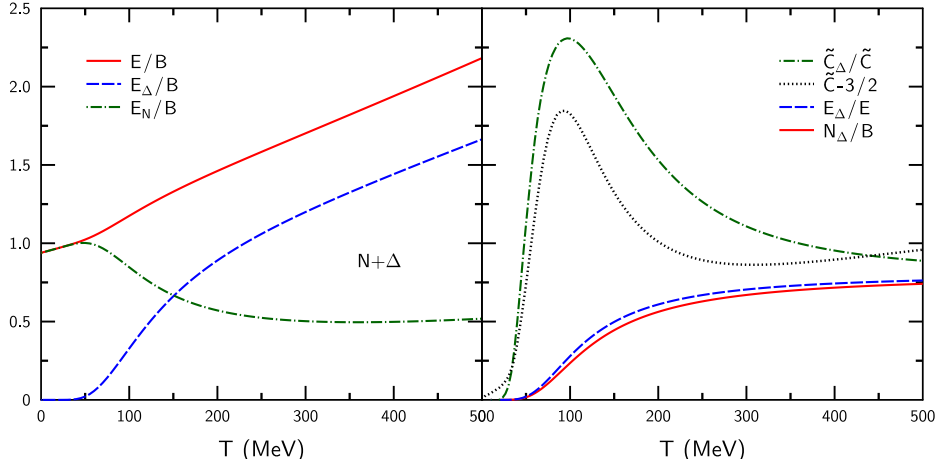}}
        \caption{(Color online)
      Left panel: the average energy per baryon of $N+\Delta$ matter $E/B$
      and partial contributions from nucleons and $\Delta$'s as functions of temperature (all energies are given
      in~GeV)\hsp . Right panel: the dashed, solid and dash-dotted curves show, respectively, relative contributions
      of resonances to energy, baryon charge and heat capacity. The dotted line shows temperature dependence of the
      total heat capacity per baryon minus 3/2.}
        \label{fig7}
        \end{figure*}
Using~\re{chemp} one obtains the formula for the free energy density
\bel{fed1}
f=\mu_B\hsp n_B-P=n_BT\left\{\ln{\frac{n_B}{\phi_N+\phi_\Delta}}-1+
\int\limits_0^{~n_B}\hsp\frac{\ds dn_1}{\ds n_1}\left[Z(\hsp n_1)-1\right]\right\}.
\ee
This leads to the following equations for the energy density and isochoric heat capacity of the~\mbox{$N+\Delta$} mixture:
\begin{eqnarray}
&&\varepsilon = f-T\left(\frac{\partial f}{\partial\hsp T}\right)_{n_B}=
n_B T\left<\hspm\xi-1\right>,\label{endnd}\\
&&C_{\hsp V}=\left(\frac{\partial\hsp\varepsilon}{\partial\hsp T}\right)_{n_B}=
n_B\left[\left<x^2+3\hspm \xi\right>-\left<\xi-1\right>^2\right]\equiv n_B\hsp\widetilde{C}(T)\label{hcnd}\,.
\end{eqnarray}
Angular brackets in Eqs.~(\ref{endnd})--(\ref{hcnd}) denote averaging over the concentrations of $N$ and $\Delta$
particles. Namely, we define
\mbox{$<A>=\sum\limits_{i=N,\Delta} A_{\hsp i} w_i$} where $w_i$ is introduced in (\ref{pden1}) and $A_{\hsp i}$
is any quantity characterizing the $i$-th component. In particular, $<\xi>=\xi_N w_N+\xi_\Delta w_\Delta$ where
$\xi_{\hsp i}$ is defined in~\re{ksif}. As one can see from (\ref{endnd})--(\ref{hcnd}), the energy and heat
capacity densities are the same as in the ideal
gas of $N+\Delta$ particles~\cite{Gal79}\hsp .

     \begin{figure*}[hbt!]
          \centerline{\includegraphics[trim=0 7.5cm 0 8.5cm, clip, width=0.8\textwidth]{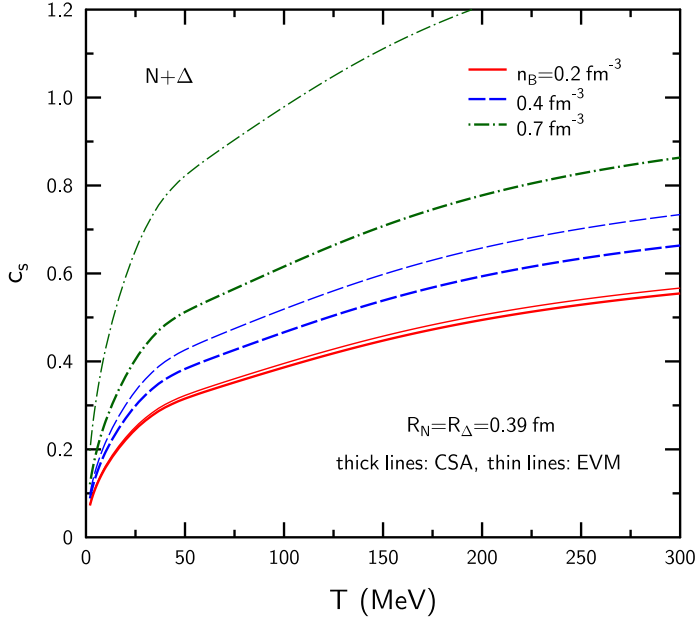}}
        \caption{(Color online)
         Sound velocity of $N+\Delta$ matter as a function of temperature for
         several values of baryon density $n_B$\hsp . Thick and thin curves give the results of CSA and EVM, respectively.}
        \label{fig8}
        \end{figure*}
The left panel of Fig.~\ref{fig7} shows the temperature dependence of the total energy per baryon~$E/B$ as
well as the partial contributions to this quantity, \mbox{$E_{\hspm i}/B=T(\xi_i-1)\hsp w_i$} ($i=N,\Delta$)\hsp .
At~fixed baryon charge \mbox{$B=N_N+N_\Delta$} the equilibrium
number of~$\Delta$'s increases with temperature (see the right panel). According to 
Fig.~\ref{fig7}, excitation
of resonances becomes \mbox{important} at~\mbox{$T\gtrsim 50~\textrm{MeV}$}. It is interesting to note that the
nucleon part of energy,~$E_N$, drops with temperature\hspm\footnote
{
This occurs because the growth of the nucleon single particle energy, $E_N/N_N=T(\xi_N-1)$\hspm ,
with raising~$T$ is compensated by a stronger decrease of the number of nucleons $N_N(T)=w_N(T)\hspm B$\hspm .
}
in the interval of $T$ approxi\-mately between 50 and 350~MeV.
Introducing the partial components $\widetilde{C}_i=B^{-1}dE_i/d\hsp T$ of the total heat
capacity~$\widetilde{C}$ we conclude that the nucleon contribution,
\mbox{$\widetilde{C}_N=\widetilde{C}-\widetilde{C}_\Delta$}, is negative in the above mentioned interval of~$T$.
One can see from Fig.~\ref{fig7} that $\widetilde{C}_\Delta>\widetilde{C}$ in this region.

     \begin{figure*}[hbt!]
          \centerline{\includegraphics[trim=0 7.5cm 0 8.5cm, clip, width=0.8\textwidth]{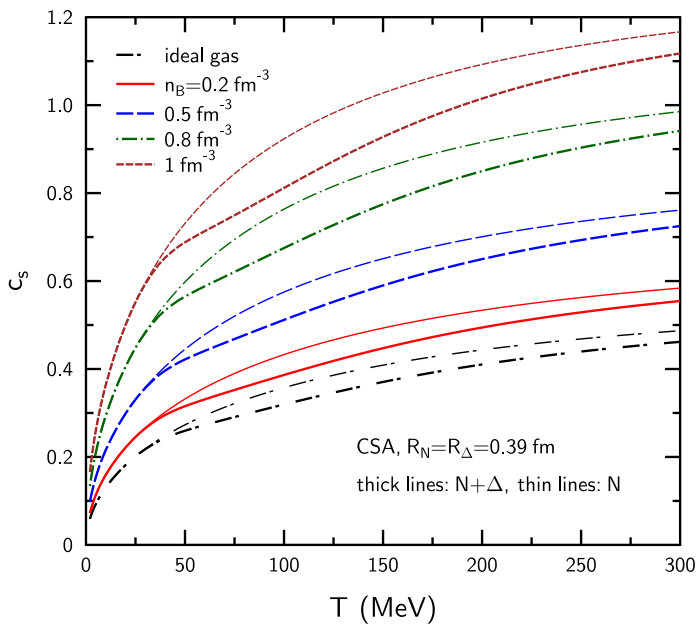}}
        \caption{(Color online)
         Sound velocities of baryon matter as functions of temperature for different values of baryon
         density $n_B$\hspm . Thin and thick lines correspond, respectively, to the nucleon matter and to the
	  $N+\Delta$ mixture.
         All calculations are made in CSA.}
        \label{fig9}
        \end{figure*}
Now we calculate the sound velocity of equilibrium hadronic matter by using the general formula
\bel{adsv3}
c_s^{\hsp 2}=\frac{1}{\varepsilon+P}\left[\hsp n_B\left(\frac{\partial\hsp P}
{\partial\hspace*{1.1pt}\mbox{$n_B$}}\right)_T+\frac{T}{C_{\hsp V}}
\left(\frac{\partial\hsp P}{\partial\hsp T}\right)_{n_B}^2\right].
\ee
This expression is derived in the Appendix using~\re{adsv}
and basic thermodynamic iden\-tities\hsp\footnote
{
A non-relativistic version of (\ref{adsv3}) has been suggested in~Ref.~\cite{Ros99}\hsp .
}\hspace*{-1pt}.
Note that calculating $c_s$ from~\re{adsv3} does not require an explicit form of the Poisson adiabat:
one should know only $P$ and $\varepsilon$ as well as their partial derivatives with respect to~$T$
and~$n_B$\hsp . We would like to stress that \re{adsv3} is applicable for any form of short-range
interaction, for any number of hadronic species (including antibaryons and strange particles) and
can be used even in the case of quantum
statistics.

For a given EoS\hsp , one can use \re{adsv3} to check constraints imposed by
the \mbox{causality} condition~$c_s\leqslant 1$\hspm . For example, for the
polytropic EoS $P=\alpha\hsp n_B^{\hsp\gamma}$ at zero temperature, \re{adsv3} predicts
that~\mbox{$c_s^{\,2}=P^{\hsp\prime}(n_B)\left[\int\limits_0^{n_B}
\frac{\ds dn}{\ds n}\hsp P^{\hsp\prime}(n)\right]^{-1}\hspace*{-5pt}=\gamma-1$}.
Therefore, in this case the~parameter $\gamma$ should satisfy the condition\hsp\footnote
{
The causal limit for the polytropic EoS was first considered in Ref.~\cite{Zel62}.
}
$1\leqslant\gamma\leqslant 2$\hspm .

One can calculate the sound velocity of the $N+\Delta$ mixture using (\ref{adsv3}) and formulae
for~$P,\varepsilon,C_{\hsp V}$ derived in this section. We arrive at the following result
\bel{svnd1}
c_s^{\hsp 2}=\frac{(n_BZ)^{\hsp\prime}+Z^{\hsp 2}\widetilde{C}^{\hsp -1}}{<\xi>+\,Z-1}\,,
\ee
where prime denotes the derivative with respect to $n_B$. Note that this formula can also be obtained
from~\re{adsvn} if one replaces $n,\xi$ by $n_B,<\xi>$ and uses, instead of (\ref{hcap}), the expression (\ref{hcnd})
for $\widetilde{C}$\hspm .

Figure~\ref{fig8} shows the results of $c_s$--\,calculation in the EVM and CSA. We choose the hadron hard-core
radius $R=0.39~\textrm{fm}$\hspm . Again one can see that the CSA predicts smaller sound velocities than the EVM.
As compared to the EVM, the sound velocity in the CSA  increases with~$n_B$ much slower, but the temperature dependence
is rather similar. In Fig.~\ref{fig9} we compare the sound velocities of the nucleonic and $N+\Delta$ matter.
Both calculations are made in the CSA. The ideal gas results are obtained by taking the
limit~$n_B\to 0$\hspm . One can see that inclusion of resonances leads to a noticeable reduction of sound
velocities at $T\gtrsim 50~\textrm{MeV}$. For realistic temperatures $T\lesssim 200~\textrm{MeV}$, superluminal
values $c_s>1$ appear only at baryon densities $n_B\gtrsim 1~\textrm{fm}^{-3}$.

The $N+\Delta$ mixture considered so far can not be regarded as a realistic system at
high temperatures. In this case mesons will be copiously produced due to inelastic collisions of baryons and decays
of resonances. To take these processes into account, below we study the EoS of a three--component $\pi+N+\Delta$ mixture.
In this study we again assume that hadrons interact via HS\hspm I and neglect possible differences in baryonic radii, i.e.
we take \mbox{$R_N=R_\Delta=R$}\hspm . To estimate the sensitivity to the pion size, we consider two
limiting cases. First, we assume equal radii for all species, i.e. choose $R_{\,\pi}=R$, and then we
investigate the~$\pi+N+\Delta$ mixture with point--like pions ($R_{\,\pi}=0$)\hspm .

\subsection{The $\pi+N+\Delta$ matter (same sizes of hadrons)}

\begin{figure*}[hbt!]
          \centerline{\includegraphics[trim=0 7cm 0 9cm, clip, width=0.8\textwidth]{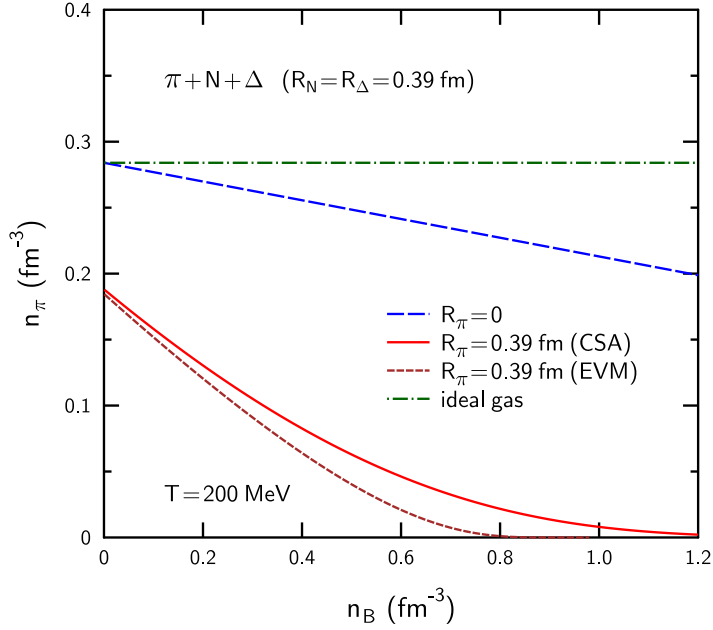}}
        \caption{(Color online)
         Equilibrium pion density in \mbox{$\pi+N+\Delta$} mixture as a function of baryon density
	  for~$T=200~\textrm{MeV}$. The solid and short-dashed curves are calculated within CSA and EVM assuming
	  equal sizes of hadrons. The long-dashed line is obtained in the limit of point--like pions. The dashed-dotted
	  curve corresponds to the ideal gas.}
        \label{fig10}
        \end{figure*}

In this section we consider the $\pi+N+\Delta$ mixture assuming equal sizes of all hadrons.
In this case one can find shifts of thermodynamic functions in the same way
as in Sec.~\ref{chsi} for a one--component system. The only difference is that instead of particle density $n$
one should substitute the total density of hadrons~$n_{\hspm\pi}+n_B$\hspm . As a result, we obtain the relations
\begin{eqnarray}
&&P=nTZ(n),~~~n=n_{\hspm\pi}+n_B\,,\label{pres3c}\\
&&\mu_{\hspm i}=T\left[\ln{\frac{n_{\hspm i}}{\phi_{\hspm i}(T)}}+\psi(n)\right],\label{chep3}
\end{eqnarray}
for pressure and chemical potentials of particle species~$i=\pi,N,\Delta$\hspm .
Using further~\re{cheq}, one gets the equations for equilibrium pion density
\bel{pid2}
n_{\hspm\pi}=\phi_{\hspm\pi} (T)\,e^{-\hsp\ds\psi\hspm (n_{\hspm\pi}+n_B)},
\ee
and for baryon chemical potential
\bel{mube}
\mu_B=T\left[\ln{\frac{n_B}{\phi_N+\phi_\Delta}}+\psi(n_{\hspm\pi}+n_B)\right].
\ee
Solving (\ref{pid2}) with respect to $n_\pi$ and substituting the result
into~(\ref{pres3c}) gives the equilibrium pressure $P=P\hsp (T,n_B)$ of the considered mixture.
Similarly to~Sec.~\ref{ndm}, one can show that equilibrium fractions
$n_N/n_B$ and $n_\Delta/n_B$ are the same as in the ideal~$N+\Delta$ gas (see~\re{pden1})\hspm .
According to \re{pid2}, interaction with baryons leads to suppression of pion density as compared to pure pion
gas ($n_B=0$)\hsp . This is demonstrated in Fig.~\ref{fig10}
(see the solid and short-dashed curves)\hsp .

     \begin{figure*}[hbt!]
          \centerline{\includegraphics[trim=0 5.7cm 0 5cm, clip, width=1.3\textwidth]{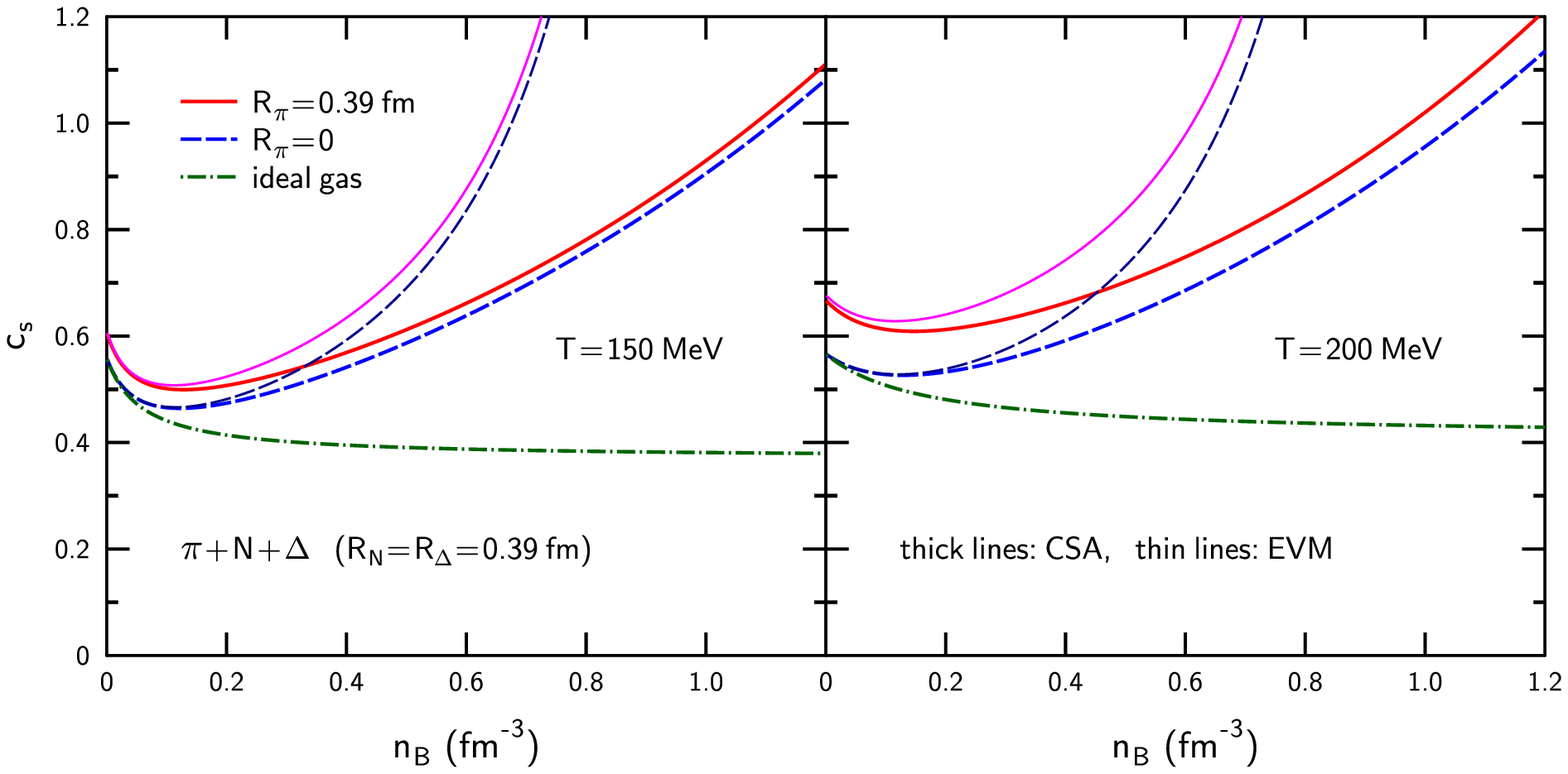}}
        \caption{(Color online)
         Sound velocity of \mbox{$\pi+N+\Delta$} matter as a function of baryon density for~$T=150~\textrm{MeV}$
	  (left panel) and $200~\textrm{MeV}$ (right panel). Thick and thin curves give, respectively, the results of~CSA
	  and EVM. The solid lines correspond to the case of equal sizes of baryons and pions. The dashed lines show the
	  results in the limit of point--like pions. The dashed-dotted curves correspond to the ideal gas.}
        \label{fig11}
        \end{figure*}
Calculating further the free energy density $f=\mu_B\hsp n_B-P$ and using Eqs. (\ref{ther1}), (\ref{pid2})--(\ref{mube})
lead to the following equations for the energy density and heat capacity
\begin{eqnarray}
&&\varepsilon =T\left[\hsp n_B\left<\hspm\xi-1\right>+n_{\hspm\pi}(\xi_{\pi}-1)\hsp\right],\label{endnd1}\\
&&C_{\hsp V}=n_B\left[\left<x^2+3\hsp\xi\right>-\left<\hsp\xi-1\right>^2\right]+
n_{\hspm\pi}\left[\hsp x^2_\pi+3\hsp\xi_\pi-(\xi_\pi-1)^{\hsp 2}(1-\chi)\hsp\right]\label{hcnd1},
\end{eqnarray}
where
$\chi=1+\left(\frac{\ds\partial\hsp n_{\hspm\pi}}
{\ds\partial\hsp n_B}\right)_T=\left[\hsp 1+n_{\hspm\pi}\psi^{\hsp\prime}(n)\hsp\right]^{-1}$
(here and below prime denotes the derivative with respect to $n$)\,.
Note that \re{endnd1} formally corresponds to the ideal $\pi+N+\Delta$ gas\hspm , but
with reduced pion density $n_{\hspm\pi}<n_{\hspm\pi}^{\hsp\rm id}=\phi_{\hsp\pi}\hspm (T)$\hsp .

Using~\re{adsv3}, one can calculate the sound velocity of the considered matter. We use the following
expressions for the derivatives of pressure
\bel{prdr}
\left(\frac{\ds\partial\hsp P}{\ds\partial\hsp n_B}\right)_T=T\chi (nZ)^{\hsp\prime},~~~
\left(\frac{\ds\partial\hsp P}{\ds\partial\hsp T}\right)_{n_B}=nZ+\chi (nZ)^{\hsp\prime}n_{\hspm\pi}(\xi_{\pi}-1)\,.
\ee
The results of $c_s$--\,calculation are shown by the solid lines in~Fig.~\ref{fig11} for two values of temperature.
The local minima of $c_s$ at \mbox{$n_B\sim 0.2~\textrm{fm}^{-3}$} appear due to a non-monotonic behavior of the total
density $n_{\hspm\pi}+n_B$ as a function of $n_B$\hsp . Again one can see that compared to EVM\hspm , the region of
superluminal sound velocities in CSA is shifted to higher baryon densities.

\subsection{The $\pi+N+\Delta$ mixture with point--like pions}

Finally we consider the limiting case of point--like pions (\mbox{$R_{\,\pi}=0$})\hsp . In accordance with~\re{pr2cs},
in this case one can represent pressure of the~\mbox{$\pi+N+\Delta$} system as
\bel{ppnd2}
P=P\hsp (T,n_\pi,n_N,n_\Delta)=T\left[\frac{n_{\hspm\pi}}{1-\eta}+n_B\hsp Z(n_B)\right],
\ee
where $\eta=v\hspm n_B$ ($v$ is the proper volume of a baryon) and \mbox{$n_B=n_N+n_\Delta$}\hsp . The compressibility
factor $Z$ describes the contribution of baryon interactions.
Below we use the parametrizations of $Z$ from Eqs.~(\ref{zevm}), (\ref{zcsa})\hsp . Substituting
\mbox{$\Delta P=P-(n_{\hspm\pi}+n_B)\hsp T$} into \re{shfe}\hsp , one can write the shift of free energy density
as follows
\bel{feds4}
\hspace*{-1mm}\Delta f=f-T\hspace*{-2pt}\sum\limits_{i=\hsp\pi,N,\Delta}\hspace*{-2pt}
n_{\hspm i}\left[\ln{\frac{n_{\hspm i}}{\phi_{\hspm i}(T)}}-1\right]
=T\left\{n_{\hspm\pi}\ln{(1-\eta)^{-1}}+n_B\int\limits_0^{~n_B}\hsp\frac{\ds dn_1}{\ds n_1}\left[Z(\hsp n_1)-1\right]\right\}.
\ee
     \begin{figure*}[hbt!]
          \centerline{\includegraphics[trim=0 7.5cm 0 9cm, clip, width=0.8\textwidth]{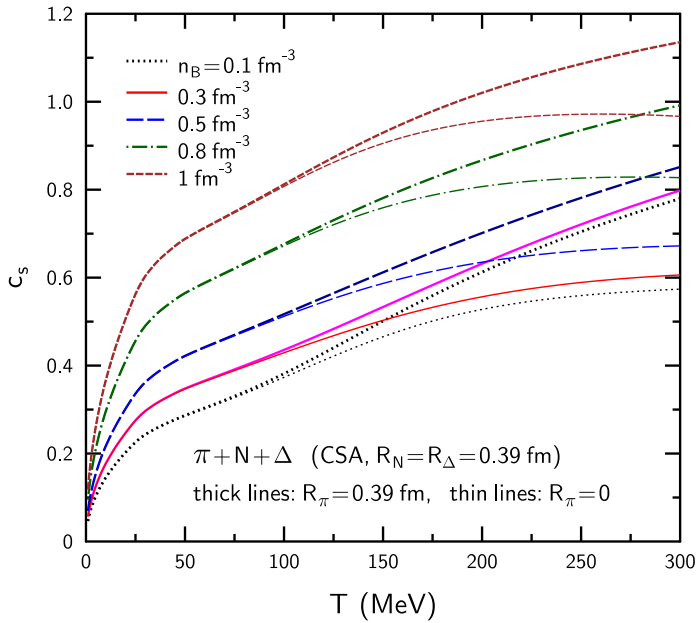}}
        \caption{(Color online)
         Sound velocity of \mbox{$\pi+N+\Delta$} matter as a function of temperature for different values of baryon
	  density $n_B$. Thick and thin lines are calculated within the CSA assuming, respectively, the pion radii
         $R_{\hsp\pi}=0.39~\textrm{fm}$ and $R_{\hsp\pi}=0$\hsp .}
        \label{fig12}
        \end{figure*}

Using further the conditions of chemical equilibrium
\bel{cche3}
\mu_{\hspm\pi}=\frac{\partial\hsp f}{\partial\hsp n_{\hspm\pi}}=0\,,~~\mu_B=
\frac{\partial\hsp f}{\partial\hsp n_{\hspm N}}=\frac{\partial\hsp f}{\partial\hsp n_{\hspm\Delta}}\,,
\ee
one gets the equations for equilibrium densities $n_{\hspm i}~(\hsp i=\pi,N,\Delta)$ as functions of $T$ and~$n_B$\hspm .
In this way we obtain the same formulae for $n_N$ and $n_\Delta$ as for $N+\Delta$ matter
(see~\re{pden1})\hsp .

Inclusion of pions modifies the baryon chemical potential as compared to the $N+\Delta$ system. It is
is given by~\re{chemp} with the additional term,  $\delta\hspm\mu_B=T\hspm\phi_{\hsp\pi}v$, in
the r.h.s. This contribution has a clear physical meaning. Indeed, to add one baryon
to the system of point-like pions, one should create a cavity of volume $v$. At fixed tempera\-ture this requires
the additional
energy (work) $\delta E=P_\pi v$, where $P_\pi=T\phi_{\hsp\pi}$ is the partial pressure of pions (see below). Therefore,
the baryon chemical potential should be shifted by the value~$\delta\hspm\mu_B=P_\pi v$. Note that this shift may be
significant even at small $n_B$\hspm .

Equilibrium values of pion density and pressure can be written as
\bel{epd3}
n_{\hspm\pi}=\phi_{\hsp\pi}(\hsp T)\hsp (1-\eta)\,,~~P=T\left[\phi_{\hsp\pi}(\hsp T)+n_B\hsp Z(n_B)\hsp\right].
\ee
The last factor in the first equality describes the reduction of volume, available to pions\hspm . A~linear
decrease of $n_{\hspm\pi}$ as a function of $n_B$ is clearly seen in Fig.~\ref{fig10}\hsp .
One can also obtain the explicit formulae for $\varepsilon$ and $C_V$. They are given
by~Eqs.~(\ref{endnd1})--(\ref{hcnd1}) after substi\-tuting~$\chi=1$\hsp .

Using further~\re{adsv3} we get the equation for sound velocity squared
\bel{adsv5}
c_s^{\hsp 2}=\frac{n_B\hsp (\hsp n_B\hspm Z)^{\hsp\prime}+C_V^{\hsp -1}
(\hsp\xi_{\pi}\phi_{\hsp\pi}+n_B\hspm Z)^{\hsp 2}}
{n_B\left[\hsp <\xi>+\hsp Z-1\hsp\right]+\phi_{\hsp\pi}\left[\,\xi_{\pi}\hsp (1-\eta)+\eta\hsp\right]}~.
\ee
At small $T$, when $\phi_{\hsp\pi}\ll 1$ one obtains \re{svnd1} for the sound velocity of the~\mbox{$N+\Delta$} mixture.
In Figs.~\mbox{\ref{fig11}--\ref{fig12}} we compare the results of $c_s$--\,calculations
for $R_{\,\pi}=0.39~\textrm{fm}$ and~\mbox{$R_{\,\pi}=0$}\hspm . As expected,
at fixed $T$ and $n_B$ the sound velocity increases with $R_{\,\pi}$\hsp .
A realistic value for~$R_\pi$ is somewhere between the two considered cases. Based on the results presented in
Fig.~\ref{fig12} we conclude that the EoS for the \mbox{$\pi+N+\Delta$} mixture remains causal up to the
baryonic densities where the deconfinement phase transition is expected.

\section{Conclusions and outlook}

In this paper we have investigated the EoS and sound velocities of one-- and multi--component
hadronic systems with HSI\hspm . It is shown that widely used excluded volume models become unrealistic
at packing  fractions exceeding about 0.2\hspm . We demonstrate that the Carna\-han-Star\-ling EoS
is much softer and can be applied at much higher densities. Moreover, the sound velocity
calculated for this EoS shows the acausal behaviour only at very high baryon densities, presumably in
the region of quark-gluon phase transition. Comparing the sound velocities in hot and dense hadronic systems
with different compositions of particles, we have studied the sensitivity
of the EoS of strongly interacting matter to the formation of pions and
baryon resonances.

In the future we are going to perform similar analysis for more realistic systems which
include heavier mesons, baryons and antibaryons. Using the Carnahan-Starling EoS and
the approach suggested in Ref.~\cite{Sat09} we plan to investigate
the sensitivity of phase diagram of strongly interacting matter to finite sizes of
hadrons. In this way one can construct a realistic EoS suitable for
hydrodynamical modeling of heavy--ion collisions. In particular,
one may perform simulations similar to Ref.~\cite{Mer11} to analyze
possible signatures of the deconfinement phase transition.

It would be interesting to extend this analysis beyond the limits of Boltzmann approximation
and include quantum--statistical effects for hadronic systems with HSI\hspm . These effects
should be certainly important for dense mater at low temperatures (e.g. in compact stars).
Some attempts in this direction have been made~\cite{Sat09,Yen97,Whe09,Ste11} within the excluded 
\mbox{volume~approach.}

\section*{APPENDIX A. General formula for sound velocity}
\setcounter{equation}{0}
\renewcommand{\theequation}{A.\arabic{equation}}

Let us assume that one knows pressure $P$ and energy density $\varepsilon$ of an equilibrated matter
as functions of temperature $T$ and the baryon density $n_B$\hsp . In this case one can easily
calculate the adiabatic sound velocity $c_s$\hsp .

From the thermodynamic relation~\cite{Lan80}
$d\hsp\varepsilon=(\varepsilon+P)\hsp d\hspm n_B/n_B+n_BTd\hspm\sigma$
one gets the expressions
\bel{dere}
\left(\frac{\partial\hsp\varepsilon}{\partial\hspace*{1.1pt}\mbox{$n_B$}}\right)_\sigma=
\frac{\varepsilon+P}{n_B}~,~~~\left(\frac{\partial\hsp\sigma}{\partial\hsp T}\right)_{n_B}=
\frac{C_{\hsp V}}{n_B T}~,
\ee
where $C_{\hsp V}=(\partial\hsp\varepsilon/\partial\hsp T)_{\hsp  n_B}$ is the isochoric heat capacity.

Using the identity $dP/n_B=\sigma d\hsp T+d\mu_B$ one arrives at the relation
\bel{dersd}
\left(\frac{\partial\hsp\sigma}{\partial\hspace*{1pt}\mbox{$n_B$}}\right)_T
=\frac{\partial}{\partial\hsp n_B}\left(\frac{1}{n_B}
\frac{\partial P}{\partial\hsp T}-
\frac{\partial\mu_B}{\partial\hsp T}\right)
=-\frac{1}{n_B^2}\left(\frac{\partial P}{\partial\hsp T}\right)_{n_B}.
\ee

From Eqs. (\ref{dere})--(\ref{dersd}) one can write the equation for the temperature
derivative in the adiabatic process
\bel{derta}
\left(\frac{\partial\hsp T}{\partial\hspace*{1pt}\mbox{$n_B$}}\right)_\sigma
=-\left(\frac{\partial\hsp\sigma}{\partial\hsp n_B}\right)_T
\left(\frac{\partial\hsp\sigma}{\partial\hspace*{1.3pt}T}\right)_{n_B}^{-1}
=\frac{T}{n_B\hsp C_{\hsp V}}\left(\frac{\partial P}{\partial\hsp T}\right)_{n_B}.
\ee
Note that this equation takes the form (\ref{adtdn1}) for classical nucleons with HSI\hspm .

Equation~(\ref{adsv}) can be represented as follows
\bel{adsv2}
c_s^{\hsp 2}=\left(\frac{\partial\hsp P}{\partial\hspace*{1.1pt}\mbox{$n_B$}}\right)_\sigma
\left(\frac{\partial\hsp\varepsilon}{\partial\hspace*{1.1pt}\mbox{$n_B$}}\right)_\sigma^{-1}
=\frac{n_B}{\varepsilon+P}\left[\left(\frac{\partial\hsp P}
{\partial\hspace*{1.1pt}\mbox{$n_B$}}\right)_T+\left(\frac{\partial\hsp P}
{\partial\hsp T}\right)_{n_B}
\left(\frac{\partial\hsp T}{\partial\hspace*{1pt}\mbox{$n_B$}}\right)_\sigma\right].
\ee
Substituting (\ref{derta}) into (\ref{adsv2}) gives the formula (\ref{adsv3})
of the main text.

\begin{acknowledgments}
The authors thank M.I. Gorenstein, A.I. Ivanytskyi and V.V. Sagun for useful discussions.
K.A.B. and I.N.M. acknowledge a financial support provided by the Helmholtz
International Center for FAIR (Germany). A~partial support
from the grant NSH--932.2014.2 is  acknowledged by I.N.M. and L.M.S.
K.A.B. acknowledges a partial support of the program
''On perspective fundamental research in high--energy and nuclear physics''
launched by the Section of Nuclear Physics of NAS of Ukraine.
\end{acknowledgments}


\end{document}